\documentclass[english,twocolumn,amsmath,superscriptaddress,amssymb,aps,pra]{revtex4-1}
\usepackage[T1]{fontenc}
\usepackage[latin9]{inputenc}
\usepackage{geometry}
\usepackage{amsmath}
\usepackage{amssymb}
\usepackage{graphicx}
\usepackage{babel}
\usepackage[colorlinks, linkcolor=blue, citecolor=blue, urlcolor=blue, breaklinks=true]{hyperref}

\newcommand{\ket}[1]{\left\vert#1\right\rangle}

\begin{document}

\title{Vibrational assisted conduction in a molecular
wire}

\author{Simon Pigeon}

\affiliation{Laboratoire Kastler Brossel, UPMC-Sorbonne Universit\'es, CNRS, ENS-PSL
Research University, Coll\`ege de France, 4 place Jussieu Case 74, F-75005
Paris, France.}

\affiliation{Centre for Theoretical Atomic, Molecular and Optical Physics, School
of Mathematics and Physics, Queen\textquoteright s University Belfast,
Belfast BT7 1NN, United Kingdom }

\author{Lorenzo Fusco}

\affiliation{Centre for Theoretical Atomic, Molecular and Optical Physics, School
of Mathematics and Physics, Queen\textquoteright s University Belfast,
Belfast BT7 1NN, United Kingdom }

\author{Gabriele De Chiara}

\affiliation{Centre for Theoretical Atomic, Molecular and Optical Physics, School
of Mathematics and Physics, Queen\textquoteright s University Belfast,
Belfast BT7 1NN, United Kingdom }

\author{Mauro Paternostro}

\affiliation{Centre for Theoretical Atomic, Molecular and Optical Physics, School
of Mathematics and Physics, Queen\textquoteright s University Belfast,
Belfast BT7 1NN, United Kingdom }

\begin{abstract}
We present a detailed study of the conduction properties of a molecular
wire where hopping processes between electronic sites are coupled to a vibrational mode of the molecule. The latter is sandwiched between two electronic
leads at finite temperatures. We show that the electro-mechanical coupling
can lead to a strong enhancement of the lead-to-lead conduction. Moreover, under suitable driving of the molecular vibrational mode, the device can act as a transistor passing sharply from enhanced conduction to short-circuit  configuration. 
\end{abstract}

\maketitle

Molecular electronics was initially dedicated to the study of transport
properties of molecules sandwiched between electronic leads \cite{Joachim2000}. It was originally conceived
as an alternative platform to silicon electronics but it has matured lately
as a rich and promising field of research beyond its initial scope \cite{Aradhya2013,Venkaratnam2006}.
Numerous theoretical works have been dedicated to enlighten the behavior
of such metal-molecule-metal junction depending on the microscopic
description of the molecule. Most of them have focused
on models including solely connected electronic sites for building
devices such as current rectifiers \cite{Aviram1974} or thermal
transistors \cite{Joulain2016} for example. Rich features were predicted
 such as stochastic pumps \cite{Sinitsyn2007,Ren2010}
or laser-induced phase-controllable transport \cite{Franco2007} to
mention only some recent results. Based on the rich nature of the molecule
under study, more recently an additional ingredient was added to the
microscopic description of these devices: the coupling of the conduction
mechanism to the vibrational degrees of freedom of the molecule, which
is expected to play a crucial role on the transport properties of these
systems \cite{Tretiak2002,Gambetta2006}. This configuration was first
considered in order to cool (through side-band cooling) the mechanical/vibrational
degree of freedom of the molecule \cite{Zippilli2009,Santandrea2011}.
The study of the impact of this coupling on transport was also considered,
showing negative differential conductance \cite{Boese2001,Walter2013,Hartle2011,Zazunov2006}
or other vibration-assisted transport phenomena \cite{Yar2011,Chen2005,Paaske2005,Koch2006,Egger2008}.
However, this extended literature on the subject is only, to our knowledge,
based on Anderson-Holstein-like models.

The Anderson-Holstein model consists of electronic sites sandwiched between
two electronic leads while the vibrational mode is coupled to an electronic
eigenstate, raising or lowering its energy depending on the vibrational
state \cite{Galperin2007}. In this paper we consider a system where the molecular vibration
is not coupled to an electronic eigenstate but to the hopping
mechanism taking place between two microscopic electronic sites. Consequently,
as it will be detailed, by controlling the vibration properties of the molecule, we can tune
the flux of electrons passing through the device. This allows for an enhancement of the conduction even when no voltage bias is applied between the electronic leads. Moreover, the process that we highlight makes possible a switching mechanism in which the flowing current  is deterministically turned on and off.

The remainder of this paper is organized as follows: Sec.~\ref{sec:description} introduces
the system and the working conditions considered throughout our work. In Sec.~\ref{sub:Vibrational-mode-damping}, we adiabatically eliminate the vibrational degree of freedom to end up with an effective dynamics
for the conducting part of the system. This analysis is then complemented by  
tracing out the degrees of freedom of the leads (Sec.~\ref{sub:Coupling-to-electronic}).
Sec.~\ref{sec:Wire-dynamic} is dedicated to the
dynamics and related exchange statistics with the electronic leads.
We detail the unraveling approach used to access the statistics of
exchange taking place between the system and one of the electronic
leads. 
We then pass to a systematic
study of the mean current flowing through
the device as a function of the electronic-lead configuration. For such a study, we focus on the low-temperature regime. We then switch to the study of more realistic conditions. We illustrate the conduction enhancement and demonstrate the transistor regime (Sec. \ref{sec:Conduction-enhancement}). In Sec. \ref{sub:conclusion} we summarise our results and comment on possible directions along which our study can be furthered.

\section{Description of the physical system\label{sec:description}}

\begin{figure}
\centering{}\includegraphics[width=0.45\textwidth]{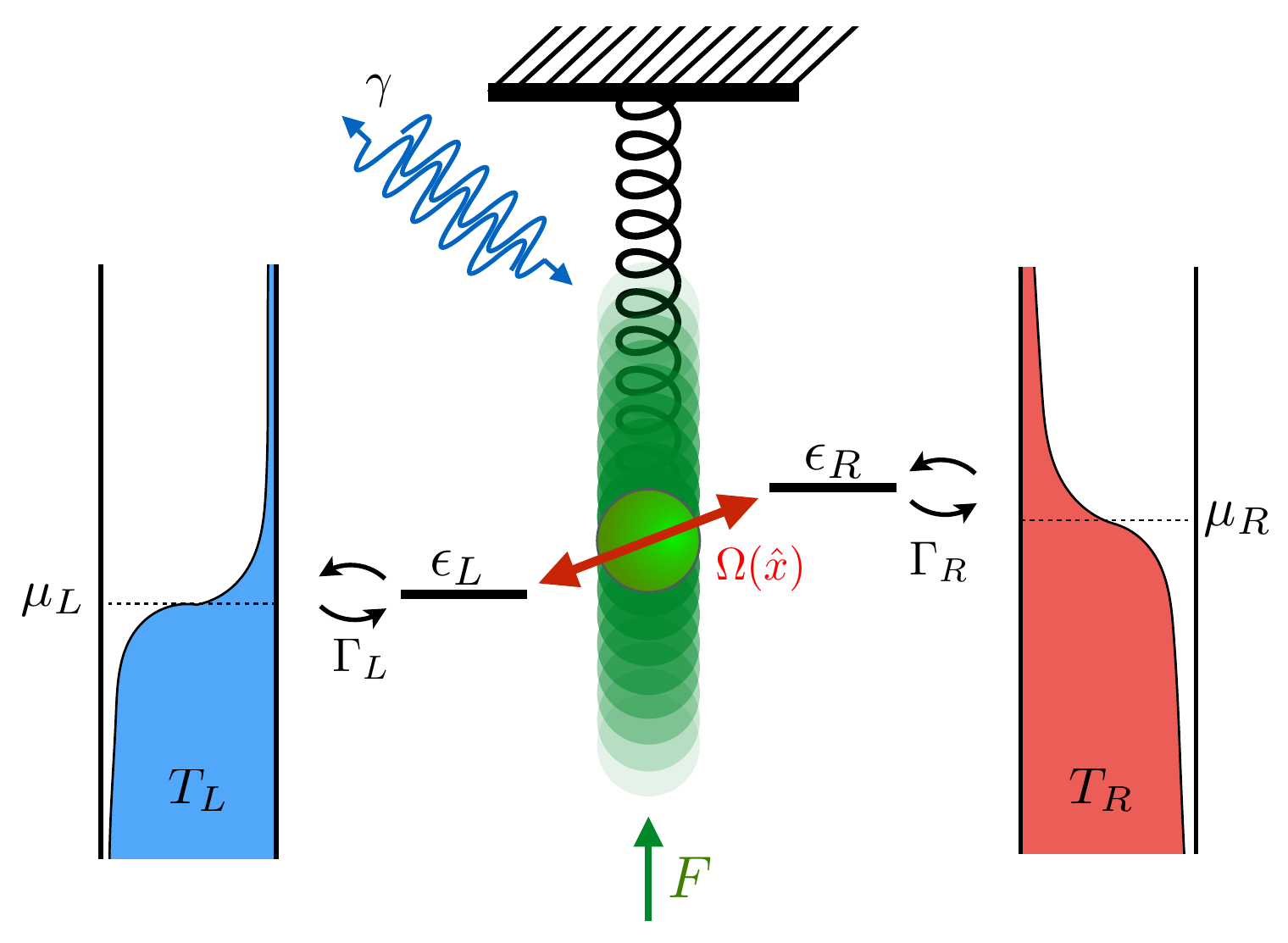}\caption{Schematics of the system representing the molecular wire connected
to two electronic leads at given temperature $T_{R}$ and $T_{L}$.
The vibrational mode is represented by the harmonic oscillator
in-between the two electronic sites and we define $\Omega(\hat{x})$ the oscillator-controlled hopping between $\epsilon_L$ and $\epsilon_R$.\label{fig:schematic-1}}
\end{figure}

As illustrated in Fig.~\ref{fig:schematic-1}, the system being considered consists of two coupled parts: {\bf (i)}
an electronic wire and {\bf (ii)} a vibrational mode. Here, we first focus on the
Hamiltonian part of the dynamics of each subsystem and their coupling. Our model aims at capturing the salient features of the  energy of a molecular wire. 
Such system consists of a three-modular molecular junction encompassing a left ($L$), right ($R$) and central ($C$) region. The $L$ and $R$ part of the junction are connected, through leads, to an external potential, thus putting the junction out of equilibrium and allowing for the circulation of an electronic current. We neglect any fluctuation of the electronic site energies. The conductance of the junction depends on the conformation of the molecule itself: by putting part $C$ out of the plane containing both $L$ and $R$ (which we assume to be coplanar), the resistance offered by the junction to the current can be varied in light of the modifications induced to the overlap of the electronic wave-functions that is responsible for the conduction. We neglect molecular reorganisation induced by electron hopping.

In what follows, we will focus on the case of small fluctuations around an otherwise stable molecular conformation (i.e. a stable relative angle between the central and peripheral parts of the molecular junction). The relative angle between the in-plane and out-of-plane parts of the junction would thus oscillate around the macroscopically stable configuration, such oscillations being treated quantum mechanically. 
A minimal model that  is able to capture the essential conformational influences over electric conductance is as follows. We consider two single-occupation sites whose energy is given by the on-site Hamiltonian 
\begin{equation}
\hat{H}_{\text{w}}=\epsilon_{L}\hat{s}_{L}^{\dagger}\hat{s}_{L}+\epsilon_{R}\hat{s}_{R}^{\dagger}\hat{s}_{R},\label{eq:hwire}
\end{equation}
where we have neglected the Coulomb interaction and have assumed units such that $\hbar=1$. In Eq.~\eqref{eq:hwire} $\hat{s}_{L/R}^{(\dagger)}$ is the annihilation (creation) operator
of an electron occupying the left ($L$) and right ($R$) site. For simplicity we restrict ourself to $\epsilon_{R}-\epsilon_{L}=\Delta>0$. 
We have assumed that the electrons do not hop directly between sites, an assumption invoked only to simplify our approach but that does not affect the phenomenology that will be illustrated here. 
Our model also includes a vibrational mode,  
which is described under harmonic approximation 
as 
\begin{equation}
\hat{H}_{\text{v}}=\omega_{\text{v}}\left(\hat{a}^{\dagger}\hat{a}+\frac{1}{2}\right)+\sqrt{2}F\hat x.
\end{equation}
Here, $\omega_{\text{v}}$ is the vibrational frequency, $\hat{a}$ ($\hat{a}^{\dagger}$) is the mode annihilation (creation) operator, and $\hat{x}=\left(\hat{a}+\hat{a}^{\dagger}\right)/\sqrt2$ is the associated position-like operator. We consider the case of a harmonic oscillator driven by a constant force of strength $F$, which may be applied on the molecule via a scanning tunnelling microscope (STM) cantilever \cite{Hugel2002} or mechanical stretching \cite{Kim2011,Perrin2013}, inducing a shift in the equilibrium position of the vibrational mode. Electronic and vibrational subsystems are coupled according to the model
\begin{equation}
\hat{H}_{\text{w-v}}=\Omega(\hat{x})(\hat{s}_{R}^{\dagger}\hat{s}_{L}+\hat{s}_{L}^{\dagger}\hat{s}_{R}).
\label{eq:coupling}
\end{equation}
Eq.~\eqref{eq:coupling} describes phonon-assisted inter-site hopping at a rate $\Omega(\hat x)$, which in turn depends on the position of the oscillator, i.e. the molecular conformation. 
The minimal scenario corresponding to the coupling in Eq.~\eqref{eq:coupling} can be realized by three $\pi$-orbitals in the edge of a triangle. Two orbitals with identical polarisation direction play the role of the electronic sites $L$ and $R$ respectively. The third one, inversely polarised, moves normally to the direction $L-R$ and plays the role of the vibrational mode. Through this motion, the third orbital will get closer/further from the conducting electrostatic cloud formed by the co-polarised orbital $L$ and $R$. This will reduce or increase their overlap and, as a consequence, suppress or enhance the conduction.
The total Hamiltonian of the system is thus 
\begin{equation}
\hat{H}=\hat{H}_{\text{w}}+\hat{H}_{\text{w-v}}+\hat{H}_{\text{v}}.
\end{equation}
The assumption of small oscillations around a stable configuration justifies a series expansion of the hopping rate as $\Omega(\hat{x})\approx\Omega_{0}+\Omega_{1}\hat{x}$. This gives rise to a standard hopping mechanism connecting the two sites being considered, and a phonon-assisted one, occurring at rate $\Omega_1$, that depends explicitly on $\hat x$. 

\subsection{Vibrational mode damping and adiabatically eliminated model \label{sub:Vibrational-mode-damping}}

Thermal excitations might induce oscillations around the stable molecular conformation. We thus consider a thermal reservoir coupled to the vibrational mode. This leads to the dynamical model 
\begin{equation}
\partial_{t}\hat{\rho}_{\text{m}}=-i[\hat{H}_{\text{v}},\hat{\rho}_{\text{m}}]+\mathcal{L}_{\text{v}}[\hat{\rho}_{\text{m}}],
\end{equation}
where $\hat{\rho}_{\text{m}}$ is the density matrix of the system and we have introduced the Lindblad dissipator 
\begin{eqnarray}
\mathcal{L}_{\text{v}}[\bullet]=\frac{\gamma\bar{n}}{2} D[\hat a^\dag,\bullet]+\frac{\gamma(\bar{n}+1)}{2} D[\hat a,\bullet],
\label{eq:ldiss-1}
\end{eqnarray}
where $\gamma$ is the coupling strength to the bosonic bath, $\bar{n}$ is the mean number of excitations in the bath (related to the bath's temperature by $\bar{n}=(\exp\left[\hbar\omega_{\text{v}}/k_{B}T_{\text{v}}\right]-1)^{-1}$), and $D[\hat O,\bullet]=(2\hat O\bullet\hat{O}^\dag-\{ \hat{O}^\dag\hat{O},\bullet\})$.
Considering Heisenberg picture, the unitary part of the dynamics corresponds to $\partial_t \hat{a} = i [\hat{H}_{\text{v}},\hat{a} ]$. Adding the dissipation and stochastic contribution, we derive the equation of motion of the annihilation operator of the vibrational mode :
\begin{equation}
\partial_{t}\hat{a}=-i\omega_{\text{v}}\hat{a}-i\frac{\Omega_{1}}{\sqrt{2}}\left(\hat{s}_{L}^{\dagger}\hat{s}_{R}+\hat{s}_{R}^{\dagger}\hat{s}_{L}\right)-\frac{\gamma}{2}\hat{a}+\hat{F}
\label{eq:cdcdc},
\end{equation}
with $\hat{F}=F+\sqrt{\gamma\bar{n}}\hat{\xi}$ where the second term represent the thermal fluctuations such as $\langle\hat{\xi}(t)\hat{\xi}^\dagger(t')\rangle=\delta(t-t')$ \cite{Gardiner2004}. 

We assume that the oscillatory mode reaches its steady-state in a time much shorter than the characteristic time of the evolution of the system, so that we can advocate for the validity of the adiabatic approximation, according to which the state of the oscillator can be assumed to be stationary and unaffected by the coupling to the electronic wire. Its degrees of freedom can thus be traced out to seek for an effective reduced dynamics of the electronic system.   
From Eq. \eqref{eq:cdcdc}, the steady-state position of the oscillator is 
\begin{equation}
\hat{x}_{st}=-\frac{1}{2} \frac{\Omega_{1}\omega_{\text{v}}}{\omega_{\text{v}}^{2}+\gamma^{2}/4}\left(\hat{s}_{L}^{\dagger}\hat{s}_{R}+\hat{s}_{R}^{\dagger}\hat{s}_{L}+\frac{\gamma\sqrt{2}}{\omega_\text{v}\Omega_1}\hat{F}\right)\label{eq:cdcd}.
\end{equation}
The Heisenberg evolution of the wire operator is 
\begin{equation}
\partial_{t}\hat{s}_{X}=-i\epsilon_{X}\hat{s}_{X}+i\left(\Omega_{0}+\Omega_{1}\hat{x}\right)\hat{s}_{Y}\hat{s}_{X}^{Z}\;,
\end{equation}
 where $\hat{s}_{X}^{Z}=[\hat{s}_{X},\hat{s}_{X}^{\dagger}]$
and with $X$ and $Y$ are either $R$ or $L$. Replacing $\hat{x}$
with its steady-state solution $\hat{x}_{st}$ in Eq.~\eqref{eq:cdcd} averaged over all oscillator trajectories, we have 
\begin{equation}
\partial_{t}\hat{s}_{X}\simeq-i\epsilon_{X}\hat{s}_{X}+i\Omega\hat{s}_{Y}\hat{s}_{X}^{Z}-i\delta\hat{s}_{Y}^{Z}\hat{s}_{X}\label{eq:l-1}\;,
\end{equation}
where $\delta=\Omega_{1}^{2}\omega_{\text{v}}/(2\sqrt{2}(\omega_{\text{v}}^{2}+\gamma^{2}/4))$
and $\Omega=\Omega_{0}-\frac{\gamma}{\omega_{\text{v}}^{2}+\gamma^{2}/4}\Omega_{1}F$.
Eq.~(\ref{eq:l-1}) can be interpreted as the Heisenberg equation for the operator $\hat{s}_X$ evolving according to the effective Hamiltonian  
\begin{equation}
\hat{H}_{\text{eff}}=\epsilon_{L}\hat{s}_{L}^{\dagger}\hat{s}_{L}+\epsilon_{R}\hat{s}_{R}^{\dagger}\hat{s}_{R}+\Omega(\hat{s}_{R}^{\dagger}\hat{s}_{L}+\hat{s}_{L}^{\dagger}\hat{s}_{R})-\delta\hat{s}_{R}^{Z}\hat{s}_{L}^{Z},\label{eq:heff}
\end{equation}
which does not contain the oscillator's degrees of freedom. 

The adiabatic elimination induces a significant
change of the coherent part of the dynamics of the electronic subsystem, but also of the incoherent part. Indeed, the reduced electronic  density matrix $\hat{\rho}$ 
evolves according to the master equation 
\begin{equation}
\partial_{t}\hat{\rho}=-i[\hat{H}_{\text{eff}},\hat{\rho}]+\mathcal{L}_{\text{eff}}[\hat{\rho}]
\end{equation}
with $\mathcal{L}_{\text{eff}}$ that describes the effective dissipation induced by the vibrational mode on the wire
subsystem
\begin{equation}
\mathcal{L}_{\text{eff}}[\bullet] = \Gamma_{\text{v}}[\hat{s}_{L}^{\dagger}\hat{s}_{R}\bullet\hat{s}_{L}^{\dagger}\hat{s}_{R}+ D[\hat{s}_{L}^{\dagger}\hat{s}_{R},\rho]+L\leftrightarrow R
]\label{eq:ldiffeff-1}\;,
\end{equation}
with $\Gamma_{\text{v}}=\gamma\delta(2\bar{n}+1)$, $\bar{n}=\exp\left[\hbar\omega_{\text{v}}/k_{B}T_{\text{v}}\right]-1)^{-1}$, and $T_{\text{v}}$ the equilibrium temperature of the oscillator. This incoherent
part of the wire subsystem dynamics encompasses two processes. The
first one, corresponding to the first term in Eq. \eqref{eq:ldiffeff-1} (and the analogous one where label $L$ is swapped with $R$),
randomly swaps the coherence between single occupancy states, while
the second term (and analogous with $L\leftrightarrow R$) allows for the incoherent hopping between both sites.
It is worth noticing that this last process leads to
an enhancement of the conduction through the wire. Moreover the strength of such an incoherent process depends
directly on the temperature of the bath attached to the vibrational
mode through the mean occupation number $\bar{n}$.
Such decoherence induced by the
molecular vibration conserves the number of excitations, stating that,
under the assumptions considered here, there is no energy exchange between
the wire and the vibronic system. 

\subsection{Coupling to electronic leads\label{sub:Coupling-to-electronic}}

We find it convenient, for the continuation of our analysis, to move to the eigenbasis of the effective Hamiltonian in Eq.~\eqref{eq:heff}, which involves non-local states \cite{Harbola2006,Gurvitz1996}, so as to get the diagonal operator
\begin{equation}
\hat{\tilde{H}}_{\text{eff}}=
\sum_{X=0}^{3}\epsilon_{X}\hat{c}_{X}^{\dagger}\hat{c}_{X}\label{eq:hwire-1-1},
\end{equation}
where $\epsilon_{X}=-\delta, (\epsilon_{L}+\epsilon_{R}+2\delta\mp\sqrt{\Delta^{2}+4\Omega^{2}})/2, \epsilon_{L}+\epsilon_{R}-\delta$
are the system energies (ordered so that $\epsilon_0<\epsilon_1<\epsilon_2<\epsilon_3$), corresponding to the non-local states $\ket{X}$.  
In what follows, we use a notation such that the states $\vert AB\rangle$ ($A, B\in\{0,1\}$) represents configurations with $A$ and $B$ electrons in the left and right site, respectively. We call these {\it local-basis} states. 

A close analysis reveals that the first and last eigenstates in
the non-local basis are identical to those of the local one, that is 
$\hat{c}_{0}^{\dagger}\hat{c}_{0}=\vert0\rangle\langle0\vert=\vert00\rangle\langle00\vert$
where no electrons are in the system and $\hat{c}_{3}^{\dagger}\hat{c}_{3}=\vert3\rangle\langle3\vert=\vert11\rangle\langle11\vert$ 
where one electron is on each sites. Notice that the Hamiltonian in Eq.~\eqref{eq:heff}
does not couple subspaces with a different total number of electrons in the sites. The intermediate
states $\ket{01}$ and $\ket{10}$ are instead coupled to give rise to the non-local states $\ket{1,2}$ as  
\begin{equation}
\begin{pmatrix}
\ket{1}\\
\ket{2}
\end{pmatrix}
=
U^{-1}\begin{pmatrix}
\ket{01}\\
\ket{10}
\end{pmatrix}
\end{equation}
with $U$ the change-of-basis matrix 
\begin{equation}
U=
\frac{2\Omega}{1-\alpha}\begin{pmatrix}
1-\alpha & 1+\alpha\\
\beta & -\beta
\end{pmatrix},
\end{equation}
where we have introduced the parameters $\alpha=\Delta/\sqrt{\Delta^{2}+4\Omega^{2}}$ and $\beta=2\Omega/\sqrt{\Delta^{2}+4\Omega^{2}}$.

Let us now connect each electronic site to a lead. The picture we have in mind can be sketched as in Fig. \ref{fig:schematic-1},
where each electronic site is individually coupled to its own fermionic
bath. The Hamiltonian of each lead is  
\begin{equation}
\hat{H}_{\nu}=\sum_{i}\epsilon_{i}^{\nu}\hat{c}_{\nu i}^{\dagger}\hat{c}_{\nu i},
\end{equation}
where $\nu\in\{L,R\}$ refers to the left or right lead and $\hat c_{\nu i}$ ($\hat c_{\nu i}^\dag$) the fermonic destruction (creation) operator for the $i^{\text{th}}$ mode in the $\nu$ bath. The coupling
between the electronic sites and each lead is of the hopping form  
\begin{equation}
\hat H_{\nu\text{-w}}=\Gamma_{\nu}\sum_{i}(\hat{c}_{\nu i}^{\dagger}\hat{s}_{\nu}+\hat{c}_{\nu i}\hat{s}_{\nu}^{\dagger})\label{eq:a}.
\end{equation}
We will now eliminate the leads' degrees of freedom. 
Writing Eq. \eqref{eq:a} in the eigenbasis of the electronic subsystem, we have 
\begin{eqnarray}
\hat{\tilde{H}}_{\nu\text{-w}} & = & \frac{\Gamma_{\nu}}{2}\sum_{a=1}^{2}\sum_{i}\Big[T_{2,a}^{\nu}\left(\hat{c}_{\nu i}^{\dagger}\hat{c}_{a}+\hat{c}_{\nu i}\hat{c}_{a}^{\dagger}\right)\nonumber \\
 &  & +T_{1,a}^{\nu}\left(\hat{c}_{\nu i}^{\dagger}\hat{c}_{3}^{\dagger}\hat{c}_{a}+\hat{c}_{\nu i}\hat{c}_{a}^{\dagger}\hat{c}_{3}\right)\Big]\label{eq:g}
\end{eqnarray}
with $\mathbf{T}^{R}=U$ and $\mathbf{T}^{L}=\begin{pmatrix}0 & 1\\
1 & 0
\end{pmatrix}U$. The first (second) line of Eq.~\eqref{eq:g} describes the transition between
the wire ground state $\vert0\rangle$ (highest energy state $\vert3\rangle$) and an intermediate state 
(either $\vert1\rangle$ or $\vert2\rangle$). As in
Refs.~\cite{Harbola2006,Gurvitz1996}, assuming the Born-Markov approximation,
we eliminate the degrees of freedoms of both leads, giving rise to two effective
dissipators $\tilde{\mathcal{L}}^{\nu}$, i.e. one per lead, 
as defined in Appendix \ref{sub:Dissipation-of-the}.
For simplicity we assume the leads to act as two similar Markovian baths with Fermi distribution at chemical potential $\mu_{\nu}$ given by
\begin{equation}
f_{\nu}(\epsilon)=\frac{1}{e^{\left(\mu_{\nu}-\epsilon\right)/k_{B}T_{\nu}}+1}~~(\nu=L,R).
\end{equation}
The number of excitations in the leads is taken to be the same (that is $n_{L}(\epsilon)=n_{R}(\epsilon)=n$)
and we choose the coupling strength $\Gamma^{L}=\Gamma^{R}=\Gamma/n\pi$. 
In this way, the only difference between the two leads consists in the respective
chemical potentials $\mu_{R}\neq\mu_{L}$. We decompose the dissipation
due to the leads, taken in the local picture as $\mathcal{L}=\mathcal{L}_{A}+\mathcal{L}_{C}$,
where the first refers to usual amplitude damping channel and the second
one to decoherence. 
{Both are explicitly defined in Appendix \ref{sub:Dissipation-of-the} [Eq.~\eqref{eq:la} and \eqref{eq:lc} respectively]}. Notice that $\mathcal{L}_{C}$ is not of a dephasing type because it changes the system
energy as well as coherences. 

Despite a rather simple system, the dynamics taking place (and especially
the dissipation acting on the system) is rich. We will now focus on
such dynamics in order to determine the exchange taking place between
the system and the leads.

\section{Wire dynamics and exchange statistics\label{sec:Wire-dynamic}}

In this Section we discuss the evolution of the electronic subsystem driven by 
 its effective dynamics and connect it to the exchange statistics with the leads, using
the thermodynamics of trajectories (Sec.~\ref{sub:Exchange-statistics}).
We then develop a systematic approach to identify the regime
maximizing the enhancement effect provided to the current crossing
the devices (Sec.~\ref{sub:Systematic}) arising from the presence of
the vibrational mode.

\subsection{Evolution of the electronic system and exchange statistics\label{sub:Exchange-statistics}}

In order to find
the stationary state of the system we restrict ourselves to the
evolution of the diagonal elements of the density matrix $\hat\rho$ in
the local basis. 
However, due to the interaction and possible hybridization of intermediate
levels, terms such as $\vert10\rangle\langle01\vert$ and $\vert01\rangle\langle10\vert$
are also crucial. The dynamics of the electronic system can be tracked by writing its density matrix in vector form as $\boldsymbol{\rho}=\big(\langle00\vert\hat\rho\vert00\rangle,\langle01\vert\hat\rho\vert01\rangle,\langle10\vert\hat\rho\vert10\rangle,\langle11\vert\hat\rho\vert11\rangle,$
$\langle01\vert\hat\rho\vert10\rangle,\langle10\vert\hat\rho\vert01\rangle\big)^{T}$, and the master equation as 
\begin{equation}
\partial_{t}\boldsymbol{\rho}=\boldsymbol{\mathcal{W}}\boldsymbol{\rho}\label{eq:dddcvf}.
\end{equation}
The form of the superoperator $\boldsymbol{\mathcal{W}}$, which is not essential for the discussions to follow, is provided in Appendix~\ref{sub:superop}. The resulting dynamics is similar to the one found in Ref.~\cite{Harbola2006}, albeit with specific features that should be stressed. Among them, the most significant is that the incoherent hopping induced by the vibrational mode acts directly on both the coherences ($\langle01\vert\hat\rho\vert10\rangle$ and $\langle10\vert\hat\rho\vert01\rangle$) of the electronic density matrix, and the occupations of the intermediate states ($\langle01\vert\hat\rho\vert01\rangle$
and $\langle10\vert\hat\rho\vert10\rangle$). 

To determine the exchange statistics taking place between the system
and the leads we now use the formalism of thermodynamics of trajectories \cite{Garrahan2010,Pigeon2014,Pigeon2016}: We define a counting process of the net exchange of excitations between the system and the right leads
such as $K:=\sum_{a=1}^{2}\sum_{b=\{0,3\}}K_{a\leftrightarrow b}^{R}-J_{a\leftrightarrow b}^{R}$,
where $K_{a\leftrightarrow b}^{\nu}$ ($J_{a\leftrightarrow b}^{R}$)
refers to an increment related to excitations leaving (entering) the
system to (from) the lead $\nu$ inducing a transition between level $a$ and $b$. 
The introduction of such process modifies the dynamics of the electronic density matrix, which now obeys
the biased master equation  
$\partial_{t}\hat{\rho}_{s}=\mathcal{W}\left[\hat\rho_{s}\right]+\mathcal{L}_{s}[\hat\rho_{s}]$
where the biasing contribution to the evolution $\mathcal{L}_{s}$
is defined in Eq. \eqref{eq:ls} and where $\hat{\rho}_{s}=\sum_{K}e^{-sK}P^{K}\hat{\rho}$ is the biased density matrix
with $P^{K}$ a projector over the subspace where the selected counting
process results in $K$ excitations being exchanged. As done previously, we consider the evolution
of the vector of relevant density matrix elements $\boldsymbol{\rho}_{s}=\big(\langle00\vert\hat\rho_{s}\vert00\rangle,\langle01\vert\hat\rho_{s}\vert01\rangle$$,\langle10\vert\hat\rho_{s}\vert10\rangle,\langle11\vert\hat\rho_{s}\vert11\rangle,$
$\langle01\vert\hat\rho_{s}\vert10\rangle,\langle10\vert\hat{\rho}_{s}\vert01\rangle\big)^{T}$, which occurs as
$\partial_{t}\boldsymbol{\rho}_{s}=\left(\boldsymbol{\mathcal{W}}+\boldsymbol{\mathcal{L}}_{s}\right)\boldsymbol{\rho}_{s}$, where $\boldsymbol{\mathcal L}_s$ is given in Appedix~\ref{sub:superop}.

One interesting feature of this system is that even the evolution of the coherences may induce changes
on the counting statistics. At this stage, we can formally access
the large deviation function $\theta(s)$ (scaled cumulant generating
function) encoding the full counting statistics for the steady-state regime, defined as
\begin{equation}
\theta(s)= \ln \text{Tr}\{\tilde{\rho}_s\} = \ln\left(\mathbf{x}^{T}\tilde{\boldsymbol{\rho}}_{s}\right)  \;,
\end{equation}
where $\hat{\tilde{\rho}}_s$ is the steady-state form of the biased density matrix
$\hat{\rho}_s(t)$, $\tilde{\boldsymbol{\rho}}_s$ its vectorized version, and $\mathbf{x}=(1,1,1,1,0,0)^{T}$.
Through the diagonalization of the propagator $\boldsymbol{\mathcal{W}}_{s}=\boldsymbol{\mathcal{W}}+\boldsymbol{\mathcal{L}}_{s}$,
one can access directly $\theta(s)$. The latter is aptly determined as the 
eigenvalue of $\boldsymbol{\mathcal{W}}_{s}$ with the longest decay time. More directly we can access also the first cumulant of the chosen exchange statistics using a
bypassing approach not requiring numerical coarse-graining, similar
to the one introduced in Ref.~\cite{Pigeon2016}. Given that $\kappa_1=-\partial_s \theta(s)\vert_{s=0} $ we have, as the first cumulant (moment), the mean current flowing
through the system 
\begin{equation}
\kappa_{1} = -\text{Tr}\left\{ \mathcal{L}_{0}'[\tilde{\rho}]\right\} =-\mathbf{x}^{T}\boldsymbol{\mathcal{L}}_{0}'\tilde{\boldsymbol{\rho}}
\end{equation}
with $\mathcal{L}_{0}'=\partial_{s}\mathcal{L}_{s}\vert_{s=0}$.

The second cumulant is related to the variance of the current
crossing the device defined as $\kappa_2=\partial_s^2\theta(s)\vert_{s=0}$ and can be determined as 
\begin{equation}
\kappa_{2} = \text{Tr}\left\{ \mathcal{L}_{0}''[\tilde{\rho}]+2\mathcal{L}_{0}'[\tilde{\rho}']\right\} = \mathbf{x}^{T}\left(\boldsymbol{\mathcal{L}}_{0}''\tilde{\boldsymbol{\rho}}+2\boldsymbol{\mathcal{L}}_{0}'\tilde{\boldsymbol{\rho}}'\right),
\end{equation}
where $\tilde{\boldsymbol{\rho}}'=\partial_{s}\tilde{\boldsymbol{\rho}}_{s}\vert_{s=0}$
is the steady-state solution of the first order biased density matrix
evolving according to 
\begin{equation}
\partial_{t}\boldsymbol{\rho}'=\boldsymbol{\mathcal{W}}\boldsymbol{\rho}'+\left(\boldsymbol{\mathcal{L}}_{0}'-\left(\mathbf{x}^{T}\boldsymbol{\mathcal{L}}_{0}'\boldsymbol{\rho}\right)\mathbf{\openone}\right)\boldsymbol{\rho},
\label{eq:vfvfv-1}
\end{equation}
and where $\mathbf{\openone}$ is the identity matrix. Notice that, by definition,
we have $\text{Tr}\left\{ \hat\rho'\right\} =\mathbf{x}^{T}\boldsymbol{\rho}'=0$.

According to this framework \cite{Pigeon2016}, after finding the steady
state solution of the density matrix $\tilde{\boldsymbol{\rho}}$
we can access the mean current, while by solving $\tilde{\boldsymbol{\rho}}$
and $\tilde{\boldsymbol{\rho}}'$ we access the
variance of the current flux. 

\subsection{Systematic study of the different regimes\label{sub:Systematic}}

\begin{figure}
\begin{centering}
\includegraphics[width=0.5\textwidth]{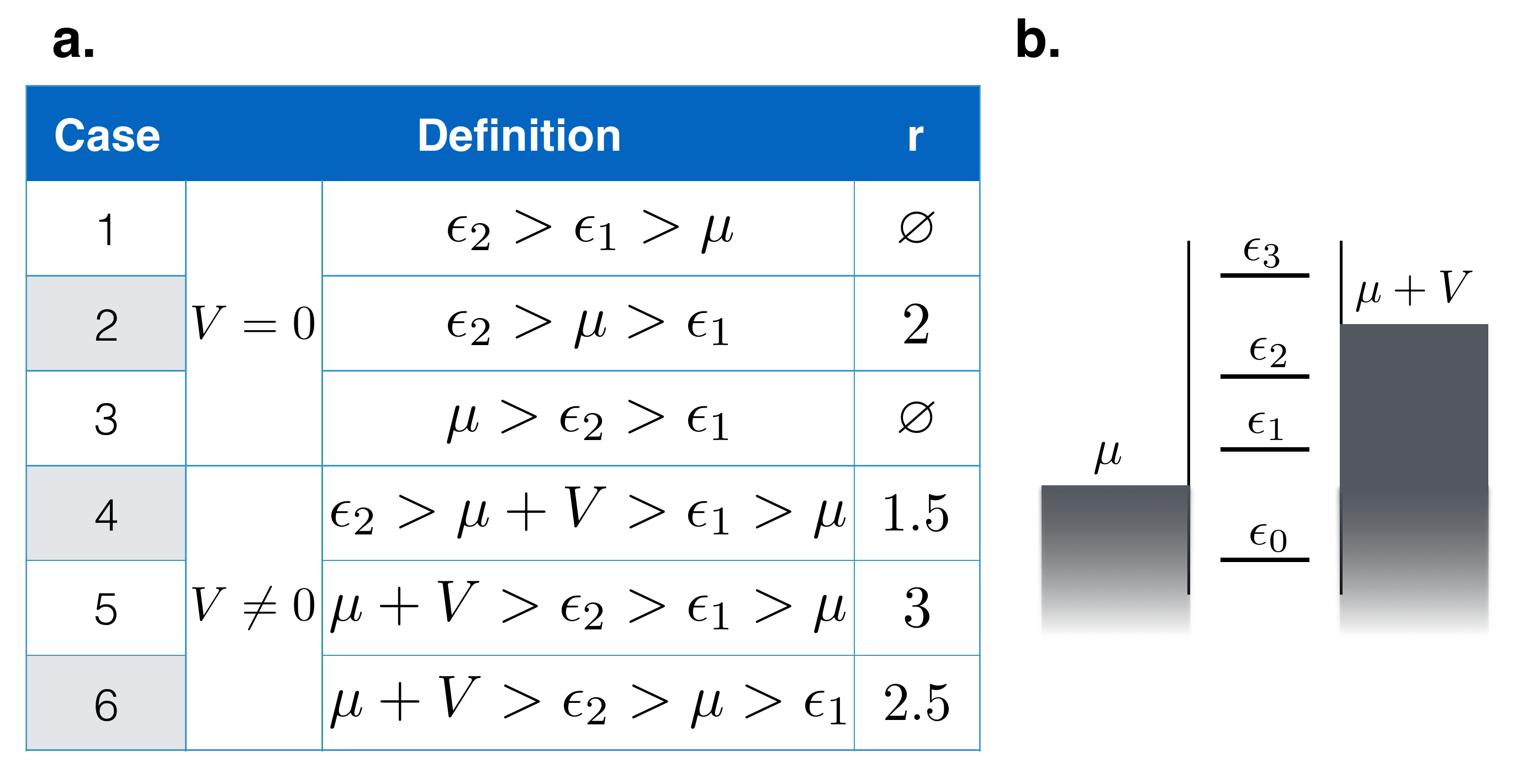}\\
\includegraphics[width=0.45\textwidth]{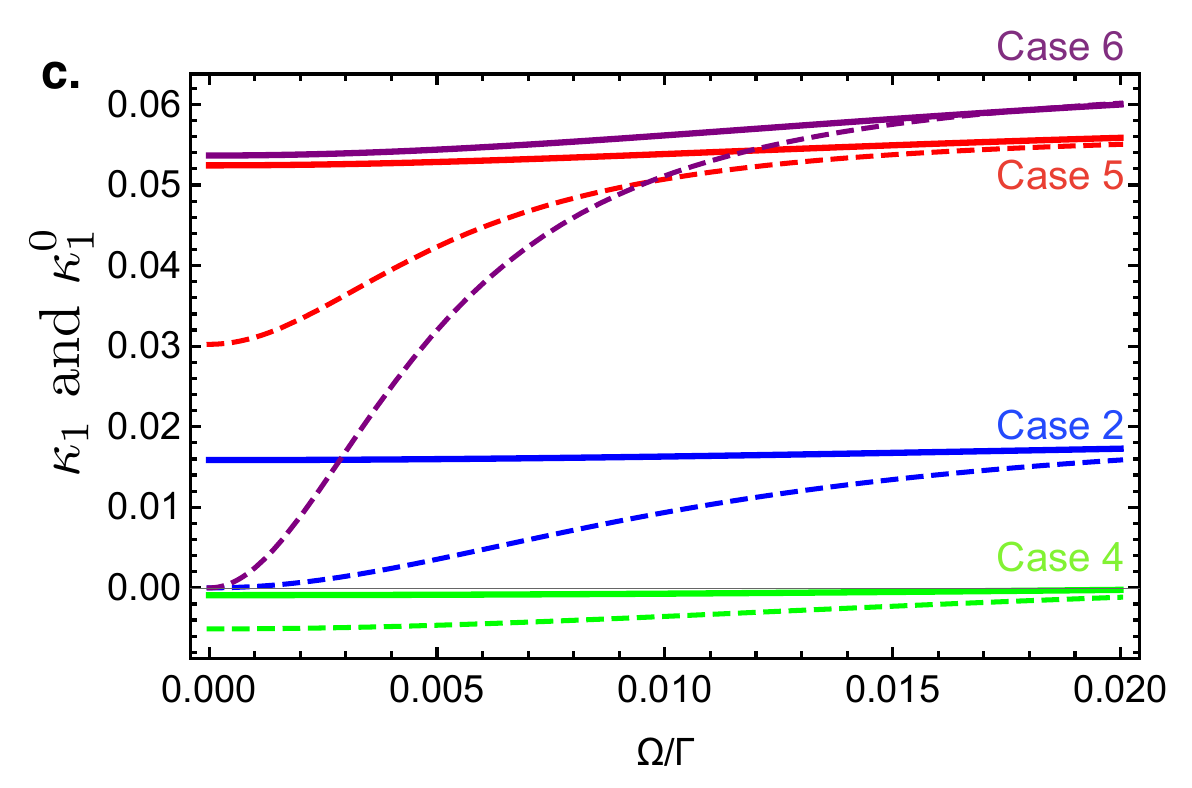}
\par\end{centering}
\caption{{\bf a.} Table summarizing the six different cases considered in the body of the manuscript. We have defined $\Omega=r\Delta$ with the value taken by $r$ reported in the third column of the table. {\bf b.}
Illustration of the energy diagram corresponding to case 5.
{\bf c.} Mean current at maximal enhancement as $\Omega=r\Delta$ as listed in table {\bf a.} under the presence of the vibrational mode ($\Gamma_{\text{v}}/\Gamma=10$) (thick full lines) and without the vibrational mode ($\Gamma_{\text{v}}/\Gamma=0$) (dashed lines). Each color refer to a specific case as indicated.}\label{fig:cases}
\end{figure}

\begin{figure}
\begin{centering}
\includegraphics[width=0.35\textwidth]{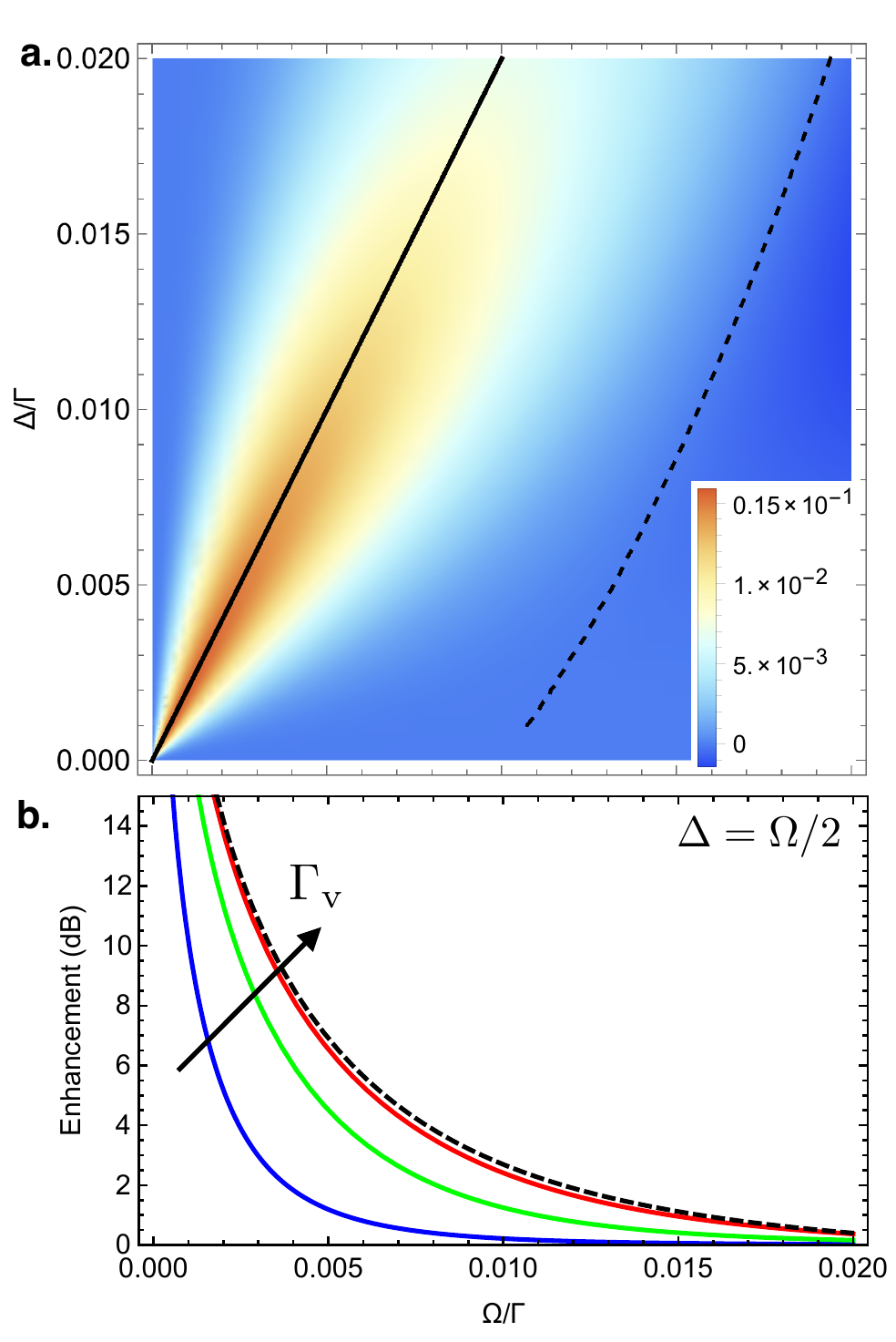}
\par\end{centering}
\caption{Enhancement induced by incoherent hopping. {\bf a.} Net enhancement $\kappa_{1}-\kappa_{1}^0$ (with $\kappa_{1}^0$ the mean current for $\Gamma_\text{v}=0$) due to the vibrational mode as a function of the energy splitting $\Delta=\epsilon_{R}-\epsilon_{L}$ (vertical axis) and the coupling strength $\Omega$ (horizontal axis).
The black full line and dashed line represent  the maximum and zero enhancement at $\Gamma_{\text{v}}/\Gamma=10$ respectively. 
{\bf b.} Maximal enhancement in dB ($10\log_{10}(\kappa_{1}/\kappa_{1}^0)$ for $\Delta=\Omega/2$) as a function of $\Omega$ for various incoherent hopping strength $\Gamma_{\text{v}}/\Gamma$ ($0.01$, $0.1$, $1$ and $\infty$). The situation presented in the figure corresponds to the case 2 in Fig. \ref{fig:cases}}.\label{fig:cases2}
\end{figure}

In order to understand the dynamics taking place we now assume that the incoherent hopping strength $\Gamma_{\text{v}}$ is independent of the other parameters. The idea here is to find out
which range of parameters gives rise to the largest enhancement due to
the presence of the vibrational mode. In order to do so we distinguish
between six different cases depending on the leads configuration,
as summarized in the table reported in Fig.~\ref{fig:cases} {\bf a.}, where we have set $\Omega=r\Delta$.
 For simplicity we consider here the case where the electronic leads
are at zero temperature, leading to the following simplification 
\begin{eqnarray}
f_{\nu}(\epsilon)
\underset{T_{\nu}\to0}{\to} \Theta(\mu_{\nu}-\epsilon),
\end{eqnarray}
where $\Theta(x)$ is the Heaviside function of argument $x$. In this condition
we find that cases 1 and 3, with no bias voltage applied between
the two leads and with their chemical potential respectively below
$\epsilon_{1}$ and above $\epsilon_{2}$, give a zero mean current. Conversely case 2, where 
$\mu$ lies between $\epsilon_{1}$ and $\epsilon_{2}$, corresponds to a non-zero
net flux from $L$ to $R$. This case is of special interest as it
allows for conduction without need of bias voltage. 
In fact, such a net flux is also present without
taking into account the molecular vibration. This unusual behaviour originates from the joint effect of coherence in the many-body states $\vert 1 \rangle$ and $\vert 2 \rangle$ and the local coupling to the electronic leads. Coherence in the delocalised states $\vert 1 \rangle$ and $\vert 2 \rangle$ allows for an electron on the left sites to be found in the right lead. This effect vanishes for large lead temperatures, and relates to observed coherent phenomena in nanoelectronics \cite{Liang2001,Darancet2009,Guedon2012}. The presence of the vibrational mode gives rise to a significant enhancement on the conduction
process, as shown in Figs.~\ref{fig:cases}.{\bf c} and \ref{fig:cases2}, where we
present the effect of incoherent hopping on the conduction 
as a function of $\Delta$, $\Omega$ and $\Gamma_{\text{v}}$ for case 2. 
As we can see in Fig.~\ref{fig:cases2}.{\bf a}, the maximum net enhancement $\kappa_{1}-\kappa_{1}^0$ (with $\kappa_{1}^0$ the mean current neglecting the vibrational mode, i.e. for $\Gamma_\text{v}=0$) occurs for $\Omega=2\Delta$. The enhancement increases as $\Omega$ and
$\Delta$ tend to zero. This comes from the fact that for $\Omega\to0$ we have clearly $\kappa_{1}\to0$ if $\Gamma_{\text{v}}=0$ (blue dashed line in Fig. \ref{fig:cases}.{\bf c}), while for $\Gamma_{\text{v}}\neq0$ and $\Omega=r\Delta$, $\kappa_{1}$ tends to be constant (blue full line in Fig. \ref{fig:cases}.{\bf c}). Consequently, in this case, for small $\Delta$ and $\Omega$, the conduction process is dominated by incoherent hopping
induced by the vibrational mode. In Fig.~\ref{fig:cases2}.{\bf b} we show how the incoherent hopping strength impacts the relative enhancement $10\log_{10}(\kappa_{1}/\kappa_{1}^0)$ at the maximal net enhancement $\Omega=2\Delta$. The enhancement diverges as $\Omega\to 0$ due to the fact that $\kappa_1^0\to0$, in this regime, as discussed. The incoherent hopping effect on the conduction quickly saturates for relatively small $\Gamma_{\text{v}}/\Gamma\approx1$.
In this case we find that the
relation $\Omega=r\Delta$ gives rise to the smallest Fano factor
($\kappa_{2}/\kappa_{1}$) depending on $\Omega$ and $\Delta$, which is smaller than 1 and converging monotonously to 1 while increasing $\Omega$ and $\Delta$ as $\kappa_1$ converges to $\kappa_1^0$, as presented in Fig.~\ref{fig:cases}.{\bf c} in blue. This indicates anti-bunching on the statistics of the net number of electrons exchanged with the right lead, corresponding to non-classical current fluctuations.

The remaining three cases all include a bias voltage (cf. Fig. \ref{fig:cases}). In all of them we have observed a similar conduction net enhancement provided by the presence of the vibrational
mode as visible in Fig.~\ref{fig:cases}.{\bf c}. The only difference with respect to case 2 and Fig.~\ref{fig:cases2}.{\bf a}
is the ratio between $\Omega$ and $\Delta$ associated to the maximal enhancement, which are as listed in Fig. \ref{fig:cases}.{\bf a}.
However, while the net enhancement 
is very similar in all such cases, the qualitative behaviours are not necessary similar. For example, case 4 presents reversed
conduction flux ($\kappa_1<0$) in some region of the parameter space (for $\Omega\ge r\Delta$) as visible in Fig~\ref{fig:cases}.{\bf c}. Cases 5 and 6, conversely, present always a positive current increasing with $\Omega$, whereas in case 2 an almost constant current is observed for $\Omega=r\Delta$. Considering the reduction of the Fano factor, its behaviour does not generally correspond to the maximal enhancement. Except for case 2, the smallest Fano factor does not coincide to $\Omega=r\Delta$ with $r$ as listed in Fig. \ref{fig:cases}.{\bf a}. The smallest Fano factor was observed in both cases 5 and 6 for $\Omega/\Delta=4$.

As the main objective of our investigation was the demonstration of conduction enhancement arising from the presence of the vibrational mode, we will not go deeper into the analysis of the six cases presented herein. It is worth noticing that for $F=0$, undriven vibrational mode, $\Omega=\Omega_0$ and $\Gamma_{\text{v}}\propto\Omega_1^2$, connecting to the microscopic description. Instead, we now focus on a more physical configuration
where the interdependence of the parameter is fully taken into account.

\section{Conduction enhancement and switching effect\label{sec:Conduction-enhancement}}

In this Section we will draw a less systematic but more practical picture of the situation under scrutiny to highlight the relevance of the effect induced by the vibrational mode on the electric conduction.
From a device perspective we will illustrate how the enhancement
 takes place under realistic conditions and how the control of the
vibrational mode can lead to a control of the electronic flux. 



\begin{figure}
\begin{centering}
\includegraphics[width=0.4\textwidth]{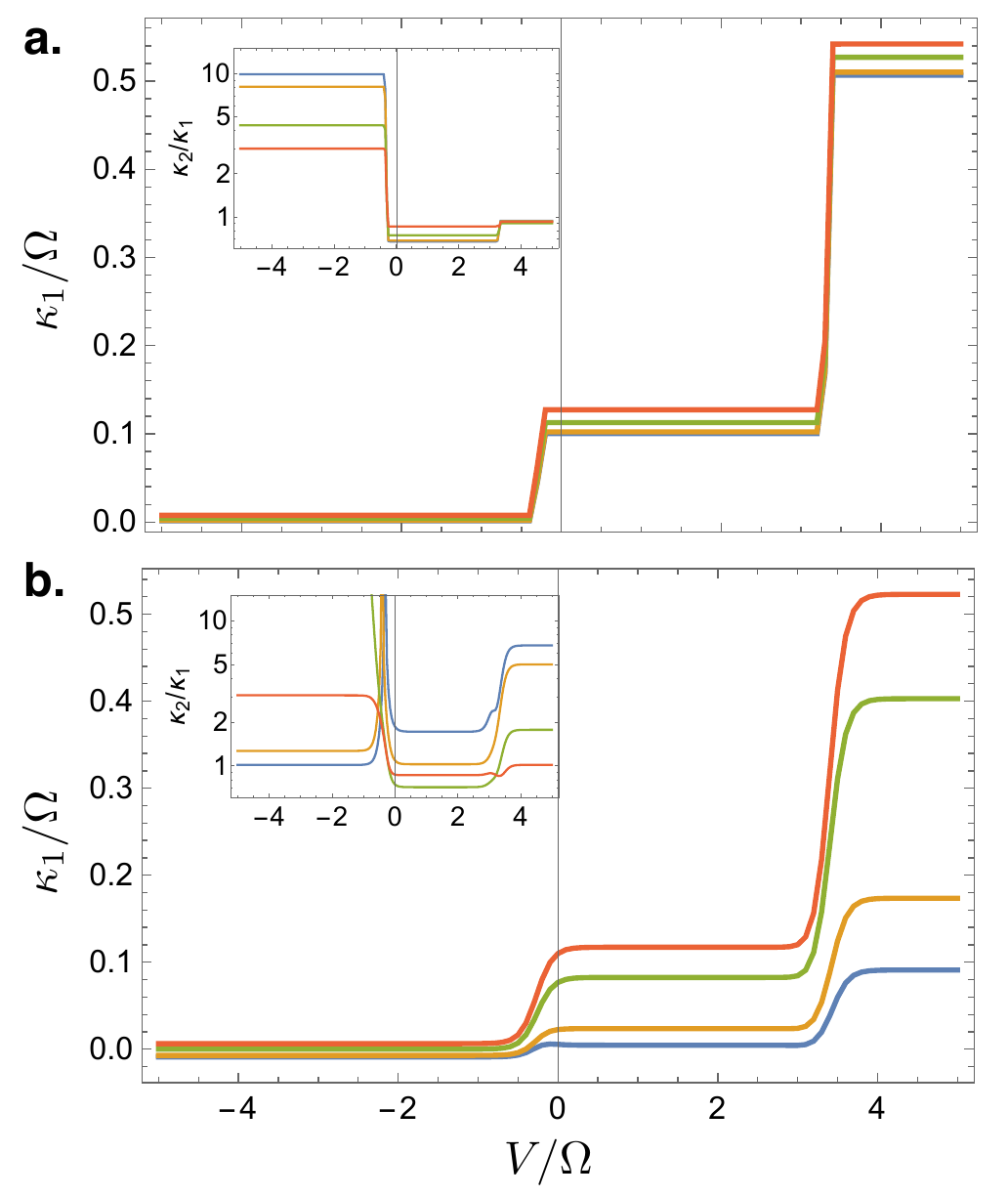}
\par\end{centering}
\caption{Net exchange as a function of the bias voltage $V$ for $\Delta/\Omega=3$
for various values of the incoherent hopping strength $\Gamma_{\text{v}}$
(respectively $0$, $0.1$, $1$ and $10$ for the blue, yellow, green
and red curves). Inset represents corresponding Fano factor. Panel {\bf a.} is for $\Omega=0.1$, while panel {\bf b.} is for $\Omega=1$. Other parameters are such as $k_{B}T_{L}=k_{B}T_{R}=0.01$,
$\delta=\epsilon_{L}=\mu=0$, and $\Gamma=1$\label{fig:Net-exchange-as}}
\end{figure}

Fig. \ref{fig:Net-exchange-as} shows the mean current $\langle I\rangle=\kappa_{1}$ passing through the wire 
as a function of the applied biased $V$, which is the most natural and accessible control parameter to adjust, and for different values of the incoherent hopping strength $\Gamma_{\text{v}}$. In the inset we represent the corresponding Fano factor. 
We notice the appearance of a plateau in the conduction in correspondence of the passage from one of the cases described in the previous Section to the next. The smoothing of the edges of those plateaux directly results from the
finite temperature used for the electronic leads. Panel {\bf a} corresponds to a case where the coherent hopping strength  $\Omega$  between
the left and right sites is one order of magnitude less
than for the lower panel {\bf b}, while keeping the ratio $\Delta/\Omega$
identical. From these two graphs we can clearly see that the enhancing effect is stronger as we decrease
$\Omega$ (and $\Delta$). The picture regarding the Fano factor, on the other hand, is less straightforward. For high $\Omega$, the changes are relatively small and monotonic with respect to $\Gamma_{\text{v}}$. For smaller $\Omega$ the changes induced by varying $\Gamma_{\text{v}}$ are more complex. The spike observed close to $V=0$ for some line, corresponds to cases where the mean current $\kappa_1$ changes sign. 
Notice that the dependence on the bias voltage is
normalized with respect to $\Omega$ to focus on the enhancing effect
provided by the vibrational mode. This renormalisation is at the origin
of a more important smoothing effect of the edge due to the temperature.
Notice also that at zero bias we have conduction taking place, and
this conduction process is not intrinsic to the presence of
the vibrational mode even if it is strongly enhanced by it.

However $\Omega$ is, in principle, not easily accessible, in particular in light of its dependence on other relevant parameters of the system. Indeed, we have 
\begin{equation}
\Omega=\Omega_{0}-\frac{\gamma}{\omega_{\text{v}}^{2}+\gamma^{2}/4}\Omega_{1}F\label{eq:cdd},
\end{equation}
where $\omega_{\text{v}}$ is the vibrational frequency, $\gamma$
its damping rate and $F$ is the driving force (if any), $\Omega_{0}$
is the bare hopping rate, independent of the molecular
vibrational properties ($0^{\text{th}}$ order expansion of $\Omega(\hat{x})$),
while $\Omega_{1}$ is the coupling strength between the oscillator
position and the electronic hopping. It is worth reminding that,
as shown previously, a key requirement to maximize the enhancement
 is to get the ratio $\Delta/\Omega$ as in Fig.~\ref{fig:cases} {\bf a.}
with $\Delta$ and $\Omega$ as small as possible. Given Eq. \eqref{eq:cdd},
fine tuning of $\Omega$ can be done through modulating the amplitude $F$ of the driving
force applied to the oscillator. The other parameter directly connected to the oscillator properties
is the incoherent hopping strength 
\begin{equation}
\Gamma_{\text{v}}=\frac{\gamma(\bar{n}+1/2)\Omega_{1}^{2}\omega_{\text{v}}^2}{\sqrt{2}(\omega_{\text{v}}^{2}+\gamma^{2}/4)^2}\label{eq:sa}.
\end{equation}
From Eq. \eqref{eq:sa} we see that raising
the temperature of the bath $T_{\text{v}}$ leads to a direct increase
of $\Gamma_{\text{v}}$. 
This dependence on the temperature is related to noise-assisted transport phenomena, which are attracting significant interest among the community \cite{Plenio2008,Viciani2015,Leon-Montiel2013}. Among others, this phenomenon was suggested to be at the origin of high efficient energy harvesting in photo-synthetic molecular complexes~\cite{Lambert2012,Levi2015}.

Consequently, by playing with a driving force $F$ applied
to the oscillator and the temperature of the corresponding bath $T_{\text{v}}$,
one can independently manipulate both $\Omega$ and
$\Gamma_{\text{v}}$. Notice also that the energy level of the system
depends on $\delta=\frac{\Omega_{1}^{2}\omega_{\text{v}}}{2\sqrt{2}(\omega_{\text{v}}^{2}+\gamma^{2}/4)}$.
This interdependence of the key parameters makes the previous systematic
approach difficult to sustain in this context. 

\begin{figure}
\begin{centering}
\includegraphics[width=0.45\textwidth]{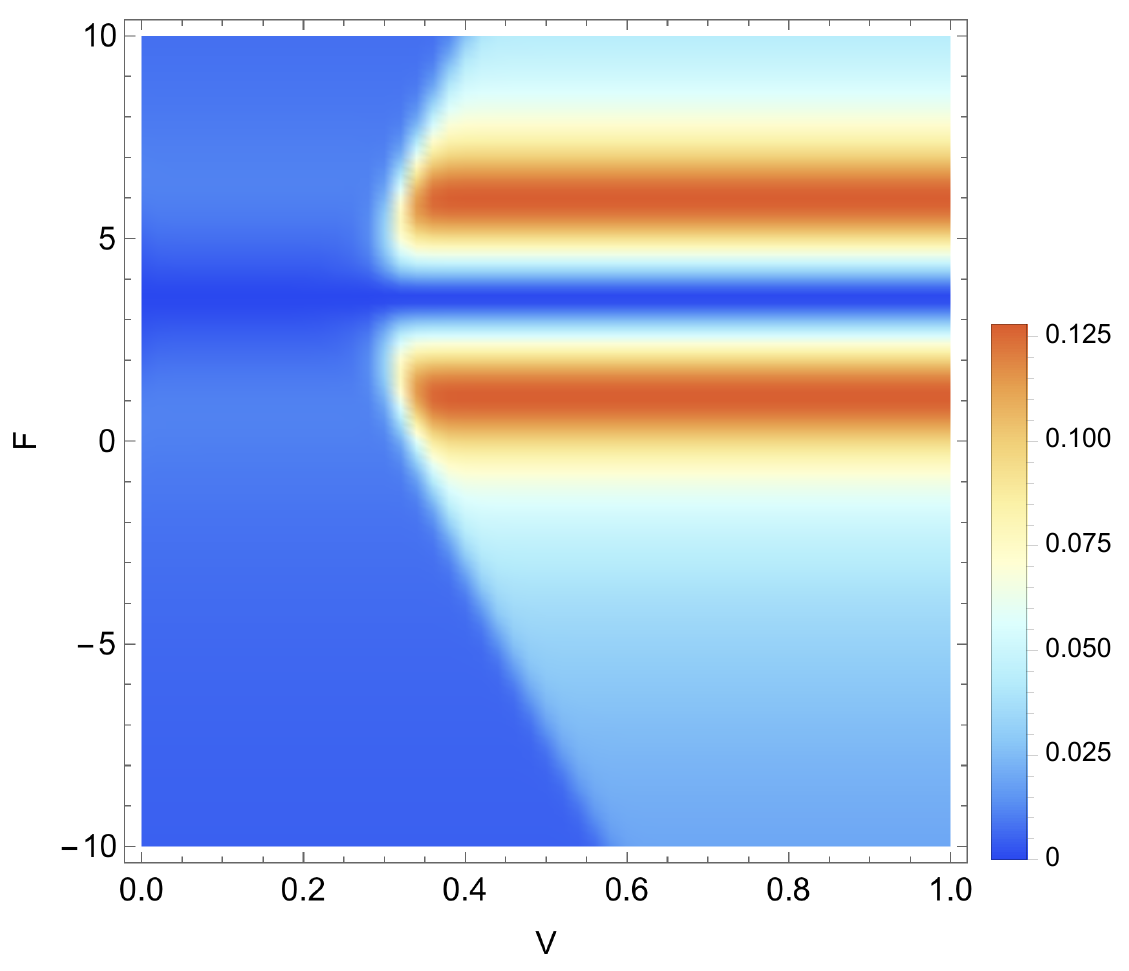}
\par\end{centering}
\caption{Mean current crossing the device ($\kappa_{1}=I$) as a function of
the voltage bias $V$ and the driving force $F$. The elected parameters
are such as $\mu=\epsilon_{L}=\Gamma=\gamma=1$, $\Omega_{1}=0.5$,
$T_{L}=T_{R}=T_{\text{v}}/100=0.01/k_{B}$, $\Delta=3\Omega_{0}=0.3$
and $\omega_{\text{v}}=0.01$ . \label{fig:real}}
\end{figure}

Nevertheless,  
Eq.~\eqref{eq:cdd}
suggests that, for the amplitude $F_{\text{crit}}=\Omega_{0}\left(\omega_{\text{v}}^{2}+\gamma^{2}/4\right)/\left(\gamma\Omega_{1}\right)$ of the driving force, the effective Hamiltonian of the wire Eq.~\eqref{eq:heff} leads to a short-circuit scenario where 
the two electronic sites become uncoupled, thus inhibiting the conduction
through the wire. In Fig.~\ref{fig:real} we plot the mean current through the device as a function of the voltage bias
$V$ and the driving force $F$. As $V$ is varied, we clearly identify the two situations illustrated in the previous Section. More specifically: for $F=0$ we have a strongly conducting regime for high $V$.
If, starting from this point in the parameter space, we increase slightly the
applied driving force toward $F_{\text{crit}}$,
we see a strong modification of the conduction, which leads to
the short-circuit regime. 
Differently from the mechanism highlighted in Ref.~\cite{Franco2007},
the switching on and off of the conduction does not require a complex
driving of the electronic sites, 
although both our scheme and Ref.~\cite{Franco2007} allow for a non-zero conduction even if no voltage
bias is applied between the electronic leads [cf. Fig.~\ref{fig:real}]. 
Consequently, beside allowing for an independent control of the
key parameters $\Gamma_{\text{v}}$ and $\Omega$ (through
the control of $F$ and $T_{\text{v}}$, respectively), in our model
the conduction can be switched on and off through a simple shift of the vibrational mode equilibrium position.

\section{Conclusion}\label{sub:conclusion}

We have presented a detailed study of the conduction properties
of a molecular wire coupled to a vibrational system via electronic hopping exchange. This description
is inspired by the idea that physically the vibrational mode does not
need to change the energetic structure of the electronic part but
can just perturb the exchange taking place on this subsystem. We showed
that the presence of the vibrational system can give rise to strong enhancement of the wire conductivity. Moreover
through the control of the vibrational properties (temperature and
position) one can accurately control the electronic flux crossing
the device. An increase of the temperature enhances the
conduction, while the control of the equilibrium position of the oscillator can switch
on and off the conduction. 

With this work, we establish how vibrational coupled hopping affects the electronic properties of a molecular wire. These crucial results pave the way to a better understanding and more complete description of electronic properties of these promising devices.


\acknowledgements
We thank the EU FP7-funded Collaborative Project TherMiQ, the John Templeton Foundation (grant number 43467), the Julian Schwinger Foundation (grant number JSF-14-7-0000), the UK EPSRC (grants EP/M003019/1 and EP/P00282X/1) and the French ANR C-FLigHT.

\bibliographystyle{plain}

\begin{thebibliography}{10}

 \bibitem{Joachim2000}
C. Joachim, J.~K. Gimzewski and A. Aviram, 
{Electronics using hybrid-molecular and mono-molecular devices},
\href{http://dx.doi.org/10.1038/35046000}{\emph{Nature} {\bf 408}, 541 (2000)}.

\bibitem{Aradhya2013}
S.~V. Aradhya and L. Venkataraman,
 {Single-molecule junctions beyond electronic transport}.
 \href{http://dx.doi.org/10.1038/nnano.2013.91}{ 
 {\em Nat. Nanotechnol.} {\bf 8}, 399 (2013)}.
 
 \bibitem{Venkaratnam2006}
L. Venkataraman, J.~E. Klare, C. Nuckolls, M.~S. Hybertsen, M.~L. Steigerward,  
{Dependence of single molecule junction conductance on molecular conformation}, \href{http://dx.doi.org/10.1038/nature05037}{{\em Nature} {\bf 442}, 904 (2006)}.

\bibitem{Aviram1974}
A. Aviram and M.~A. Ratner,
 {Molecular rectifiers}.
 \href{http://dx.doi.org/10.1016/0009-2614(74)85031-1}{ 
 {\em Chem. Phys. Lett.} {\bf 29}, 277 (1974)}.

\bibitem{Joulain2016}
K. Joulain, J. Drevillon, Y. Ezzahri and J.  Ordonez-Miranda,
 {Quantum Thermal Transistor}.
\href{http://dx.doi.org/10.1103/PhysRevLett.116.200601}{
{\em Phys. Rev. Lett.} {\bf 116}, 200601 (2016)}

\bibitem{Ren2010}
J. Ren, P. H{\"{a}}nggi and B. Li,
 {Berry-Phase-Induced Heat Pumping and Its Impact on the Fluctuation Theorem}.
\href{http://dx.doi.org/10.1103/PhysRevLett.104.170601}{ {\em Phys. Rev. Lett.} {\bf 104}, 170601 (2010)}.

\bibitem{Sinitsyn2007}
N.~A. Sinitsyn and I. Nemenman,
 {The Berry phase and the pump flux in stochastic chemical kinetics}.
\href{http://dx.doi.org/10.1209/0295-5075/77/58001}{ 
{\em EPL}, {\bf 77}, 58001 (2007)}.

\bibitem{Franco2007}
I. Franco, M. Shapiro and P. Brumer,
 {Robust ultrafast currents in molecular wires through stark shifts}.
\href{http://dx.doi.org/10.1103/PhysRevLett.99.126802}{ 
{\em Phys. Rev. Lett.} {\bf 99}, 126802 (2007)}.

\bibitem{Gambetta2006}
A.~Gambetta, C.~Manzoni, E.~Menna, M.~Meneghetti, G.~Cerullo, G.~Lanzani,
  S.~Tretiak, A.~Piryatinski, A.~Saxena, R.~L. Martin and A.~R. Bishop.
 {Real-time observation of nonlinear coherent phonon dynamics in single-walled carbon nanotubes}.
\href{http://dx.doi.org/10.1038/nphys345}{ {\em Nat. Phys.} {\bf 2}, 515 (2006)}.

\bibitem{Tretiak2002}
S.~Tretiak, A.~Saxena, R.~L. Martin and A.~R. Bishop,
 {Conformational dynamics of photoexcited conjugated molecules.}
\href{http://dx.doi.org/10.1103/PhysRevLett.89.097402}{ 
{\em Phys. Rev. Lett.} {\bf 89}, 097402 (2002)}.

\bibitem{Santandrea2011}
F.~Santandrea, L.~Y. Gorelik, R.~I. Shekhter and M.~Jonson.
 {Cooling of Nanomechanical Resonators by Thermally Activated Single-Electron Transport}.
\href{http://dx.doi.org/10.1103/PhysRevLett.106.186803}{ {\em Phys. Rev. Lett.} {\bf 106}, 186803 (2011)}.

\bibitem{Zippilli2009}
S. Zippilli, G. Morigi, and A. Bachtold.
 {Cooling Carbon Nanotubes to the Phononic Ground State with a Constant Electron Current}.
\href{http://dx.doi.org/10.1103/PhysRevLett.102.096804}{
 {\em Phys. Rev. Lett.} {\bf 102}, 096804 (2009)}.

\bibitem{Hartle2011}
R.~H{\"{a}}rtle and M.~Thoss,
 {Resonant electron transport in single-molecule junctions:
  Vibrational excitation, rectification, negative differential resistance, and local cooling}.
\href{http://dx.doi.org/10.1103/PhysRevB.83.115414}{ 
{\em Phys. Rev. B} {\bf 83}, 115414 (2011)}.

\bibitem{Walter2013}
S. Walter, B. Trauzettel and Thomas~L. Schmidt,
 {Transport properties of double quantum dots with electron-phonon coupling}.
\href{http://dx.doi.org/10.1103/PhysRevB.88.195425}{ 
{\em Phys. Rev. B} {\bf 88}, 195425 (2013)}.

\bibitem{Zazunov2006}
A. Zazunov, D. Feinberg, and T. Martin.
 {Phonon-mediated negative differential conductance in molecular quantum dots}.
\href{http://dx.doi.org/10.1103/PhysRevB.73.115405}{ 
{\em Phys. Rev. B} {\bf 73}, 115405 (2006)}.

\bibitem{Boese2001}
D.~Boese and H.~Schoeller,
 {Influence of nanomechanical properties on single-electron tunneling: A vibrating single-electron transistor}.
\href{http://dx.doi.org/10.1209/epl/i2001-00367-8}{ 
{\em EPL} {\bf 54}, 668 (2001)}.

\bibitem{Chen2005}
Z.-Z. Chen, R. L{\"{u}} and B.-F. Zhu,
 {Effects of electron-phonon interaction on nonequilibrium transport through a single-molecule transistor},
\href{http://dx.doi.org/10.1103/PhysRevB.71.165324}{ 
{\em Phys. Rev. B} {\bf 71}, 165324  (2005)}.

\bibitem{Egger2008}
R.~Egger and A.~O. Gogolin,
 {Vibration-induced correction to the current through a single molecule}.
\href{http://dx.doi.org/10.1103/PhysRevB.77.113405}{ 
{\em Phys. Rev. B} {\bf 77}, 113405 (2008)}.

\bibitem{Koch2006}
J. Koch, M. Semmelhack, F. von Oppen, and A. Nitzan,
 {Current-induced nonequilibrium vibrations in single-molecule devices}.
\href{http://dx.doi.org/10.1103/PhysRevB.73.155306}{
 {\em Phys. Rev. B} {\bf 73}, 155306 (2006)}.

\bibitem{Paaske2005}
J. Paaske and K. Flensberg, 
 {Vibrational Sidebands and the Kondo Effect in Molecular Transistors}.
\href{http://dx.doi.org/10.1103/PhysRevLett.94.176801}{ {\em Phys. Rev. Lett.} {\bf 94}, 176801 (2005)}.

\bibitem{Yar2011}
A. Yar, A. Donarini, S. Koller and M. Grifoni,
 {Dynamical symmetry breaking in vibration-assisted transport through nanostructures}.
\href{http://dx.doi.org/10.1103/PhysRevB.84.115432}{ {\em Phys. Rev. B} {\bf 84}, 115432 (2011)}.

\bibitem{Galperin2007}
M. Galperin, M.~A Ratner and A. Nitzan,
 {Molecular transport junctions: vibrational effects}.
\href{http://dx.doi.org/10.1088/0953-8984/19/10/103201}{
 {\em J. Phys. Condens. Matter} {\bf 19}, 103201 (2007)}.

\bibitem{Hugel2002} 
Hugel T, Holland N B, Cattani A, Moroder L, Seitz M and Gaub H E 2002 Single-molecule optomechanical cycle \href{https://doi.org/10.1126/science.1069856}{\emph{Science} {\bf 296} 1103}

\bibitem{Kim2011}
Y. Kim, H. Song, F. Strigl, H.-F. Pernau, T. Lee, and E. Scheer,
{Conductance and vibrational states of single-molecule junctions controlled by mechanical stretching and material variation}. 
\href{http://dx.doi.org/10.1103/PhysRevLett.106.196804}{
 {\em Phys. Rev. Lett.} {\bf 106}, 196804 (2011)}.
 
 \bibitem{Perrin2013}
M. L. Perrin, C. J. O. Verzijl, C. A. Martin, A. J. Shaikh, R. Eelkema, J. H. Van Esch, J. M. van Ruitenbeek, J. M. Thijssen, H. S. J. van der Zant, D. Duli\'c,
{Large tunable image-charge effects in single-molecule junctions}. 
\href{http://dx.doi.org/10.1038/nnano.2013.26}{
 {\em Nat. Nanotechnol.} {\bf 8}, 282 (2013)}.
 
\bibitem{Gardiner2004} C. W. Gardiner, \emph{Quantum Noise: A Handbook of Markovian and Non-Markovian Quantum Stochastic Methods with Applications to Quantum Optics} Springer Berlin Heidelberg, (2010)
 
 \bibitem{Gurvitz1996}
S.~A. Gurvitz and Ya.~S. Prager,
 {Microscopic derivation of rate equations for quantum transport}.
\href{http://dx.doi.org/10.1103/PhysRevB.53.15932}{ 
{\em Phys. Rev. B} {\bf 53}, 15932 (1996)}

\bibitem{Harbola2006}
U. Harbola, M. Esposito and S. Mukamel,
 {Quantum master equation for electron transport through quantum dots and single molecules}.
\href{http://dx.doi.org/10.1103/PhysRevB.74.235309}{ 
{\em Phys. Rev. B} {\bf 74}, 235309 (2006)}.

\bibitem{Garrahan2010}
J.~P. Garrahan and I. Lesanovsky.
 {Thermodynamics of quantum jump trajectories}.
\href{http://dx.doi.org/10.1103/PhysRevLett.104.160601}{ 
{\em Phys. Rev. Lett.} {\bf 104}, 160601 (2010)}.

 \bibitem{Pigeon2014}
S. Pigeon, L. Fusco, A. Xuereb, G. {De Chiara} and M. Paternostro,
 {Thermodynamics of trajectories of a quantum harmonic oscillator
  coupled to $N$ baths}.
\href{http://dx.doi.org/10.1103/PhysRevA.92.013844}{ {\em Phys. Rev. A} {\bf 92}, 013844 (2015)}.

\bibitem{Pigeon2016}
S. Pigeon and A. Xuereb, 
 {Thermodynamics of trajectories of open quantum systems, step by step}.
\href{http://dx.doi.org/10.1088/1742-5468/2016/06/063203}{ 
{\em J. Stat. Mech.}, 063203 (2016)}.

\bibitem{Liang2001}
W. Liang, M. Bockrath, D. Bozovic, J. H. Hafner, M. Tinkham and H. Park, 
{Fabry - Perot interference in a nanotube electron waveguide},
\href{http://dx.doi.org/10.1038/35079517}{\emph{Nature} {\bf 411}, 665 (2001)}.

\bibitem{Darancet2009}
P. Darancet, V. Olevano and D. Mayou,
{Coherent electronic transport through graphene constrictions: Subwavelength regime and optical analogy},
\href{http://dx.doi.org/10.1103/PhysRevLett.102.136803}{\emph{Phys. Rev. Lett.} {\bf 102}, 136803 (2009)}.

\bibitem{Guedon2012}
C.~M. Guédon, H. Valkenier, T. Markussen, K.~S. Thygesen, J.~C. Hummelen, S.~J. van der Molen, 
{Observation of quantum interference in molecular charge transport}, 
\href{http://dx.doi.org/10.1038/nnano.2012.37}{\emph{Nat. Nanotechnol.} {\bf 7}, 305 (2012)}.

\bibitem{Lambert2012}
N. Lambert, Y.-N. Chen, Y.-C. Cheng, C.-M. Li, G.-Y. Chen and F. Nori,
{Quantum biology}.
\href{http://dx.doi.org/10.1038/nphys2474}{ {\em Nat. Phys.} {\bf 8}, 10 (2012)}.

\bibitem{Levi2015}
F. Levi, S. Mostarda, F. Rao and F. Mintert.
{Quantum mechanics of excitation transport in photosynthetic complexes: a key issues review.}
\href{ttp://dx.doi.org/10.1088/0034-4885/78/8/082001}{ 
{\em Rep. Prog. Phys.} {\bf 78}, 082001 (2015)}

\bibitem{Leon-Montiel2013}
R.~De~J. Le{\`{o}}n-Montiel and J.~P. Torres,
 {Highly efficient noise-assisted energy transport in classical oscillator systems}.
\href{http://dx.doi.org/10.1103/PhysRevLett.110.218101}{ 
{\em Phys. Rev. Lett.} {\bf 110}, 218101 (2013)}.

\bibitem{Plenio2008}
M.~B. Plenio and S.~F. Huelga,
 {Dephasing-assisted transport: Quantum networks and biomolecules}.
\href{http://dx.doi.org/10.1088/1367-2630/10/11/113019}{ 
{\em New J. Phys.} {\bf 10}, 113019 (2008)}.

\bibitem{Viciani2015}
S. Viciani, M. Lima, M. Bellini and F. Caruso,
 {Observation of Noise-Assisted Transport in an All-Optical Cavity-Based Network}.
\href{http://dx.doi.org/10.1103/PhysRevLett.115.083601}{
 {\em Phys. Rev. Lett.} {\bf 115}, 083601 (2015)}.





\end{thebibliography}

\appendix

\subsection*{Appendix 1: Dissipation of the wire to the leads and biased contribution to the
evolution \label{sub:Dissipation-of-the}}

In this appendix we detail the calculation leading to the Lindblad
form dissipator induced by coupling the wire subsystem to fermionic
leads as explained in section \ref{sub:Coupling-to-electronic}. Based
on Refs. \cite{Harbola2006,Gurvitz1996} we determine the the
dissipator in the manybody picture such as $\tilde{\mathcal{L}}^{\nu}=\sum_{a=1}^{2}\sum_{b=\{0,3\}}\tilde{\mathcal{L}}_{a\leftrightarrow b}^{\nu}$
(with $\nu=R$ or $L$) \begin{widetext}

\begin{eqnarray}
\tilde{\mathcal{L}}_{a\leftrightarrow0}^{\nu}[\bullet] & = & \frac{\pi}{4}\Gamma^{\nu}n_{\nu}(\epsilon_{a})\vert T_{2,a}^{\nu}\vert^{2}\bigg[f_{\nu}(\epsilon_{a})\left(2\hat{c}_{a}^{\dagger}\bullet\hat{c}_{a}-\left\{ \hat{c}_{a}\hat{c}_{a}^{\dagger},\bullet\right\} \right)+(1-f_{\nu}(\epsilon_{a}))\left(2\hat{c}_{a}\bullet\hat{c}_{a}^{\dagger}-\left\{ \hat{c}_{a}^{\dagger}\hat{c}_{a},\bullet\right\} \right)\bigg]\\
\tilde{\mathcal{L}}_{a\leftrightarrow3}^{\nu}[\bullet] & = & \frac{\pi}{4}\Gamma^{\nu}n_{\nu}(\epsilon_{a})\vert T_{1,a}^{\nu}\vert^{2}\bigg[f_{\nu}(\epsilon_{a})\left(2\hat{c}_{a}^{\dagger}\hat{c}_{3}\bullet\hat{c}_{3}^{\dagger}\hat{c}_{a}-\left\{ \hat{c}_{3}^{\dagger}\hat{c}_{a}\hat{c}_{a}^{\dagger}\hat{c}_{3},\bullet\right\} \right)+(1-f_{\nu}(\epsilon_{a}))\left(2\hat{c}_{3}^{\dagger}\hat{c}_{a}\bullet\hat{c}_{a}^{\dagger}\hat{c}_{3}-\left\{ \hat{c}_{a}^{\dagger}\hat{c}_{3}\hat{c}_{3}^{\dagger}\hat{c}_{a},\bullet\right\} \right)\bigg]\;,
\end{eqnarray}
where $n_{\nu}(\epsilon)$ is the density of state in the lead $\nu$
($R$ or $L$) at a given energy and $f_{\nu}(\epsilon)=1/\left[\exp\left((\epsilon-\mu_{\nu})/k_{B}T_{\nu}\right)+1\right]$
the Fermi distribution of a given lead $\nu$ having $\mu_{\nu}$
as chemical potential. Notice that the total dissipation is related
to the leads dissipation $\tilde{\mathcal{L}}=\tilde{\mathcal{L}}^{R}+\tilde{\mathcal{L}}^{L}$
plus $\tilde{\mathcal{L}}_{\text{eff}}=U^{-1}\mathcal{L}_{\text{eff}}U$
the one induced by the vibrational mode (Eq. \eqref{eq:ldiffeff-1}).
Each dissipation induced by the leads can be decomposed in terms of
amplitude damping and coherence damping channels such as $\mathcal{L}=\mathcal{L}_{A}+\mathcal{L}_{C}$,
with for the amplitude part 
\begin{equation}
\mathcal{L}_{A}\left[\bullet\right]=\sum_{{\scriptscriptstyle XY=\{LO,LF,RO,RF\}}}\bigg[\alpha_{XY}\left(\hat{L}_{XY}\bullet\hat{L}_{XY}^{\dagger}-\frac{1}{2}\left\{ \hat{L}_{XY}^{\dagger}\hat{L}_{XY},\bullet\right\} \right)+\beta_{XY}\left(\hat{L}_{XY}^{\dagger}\bullet\hat{L}_{XY}-\frac{1}{2}\left\{ \hat{L}_{XY}\hat{L}_{XY}^{\dagger},\bullet\right\} \right)\bigg]\label{eq:la}\;,
\end{equation}
with $\hat{L}_{LO}=\vert10\rangle\langle00\vert$, $\hat{L}_{LF}=\vert10\rangle\langle11\vert$,
$\hat{L}_{RO}=\vert01\rangle\langle00\vert$ and $\hat{L}_{RF}=\vert01\rangle\langle11\vert$
and for the coupling strength we have 
\begin{eqnarray}
\alpha_{LO} & = & \frac{\Gamma}{2}\bigg[\left(1-A\right)^{2}\left(f_{L}(\epsilon_{1})B^{2}+f_{R}(\epsilon_{1})\left(1-A\right)^{2}\right)+\left(1+A\right)^{2}\left[f_{L}(\epsilon_{2})B^{2}+f_{R}(\epsilon_{2})\left(1+A\right)^{2}\right]\\
\alpha_{LF} & = & \frac{\Gamma}{2}\bigg[\left(1-A\right)^{2}\left(f_{L}(\epsilon_{1})\left(1-A\right)^{2}+f_{R}(\epsilon_{1})B^{2}\right)+\left(1+A\right)^{2}\left(f_{L}(\epsilon_{2})\left(1+A\right)^{2}+f_{R}(\epsilon_{2})B^{2}\right)\bigg]\\
\alpha_{RO} & = & \frac{\Gamma}{2}\bigg[B^{2}\left(f_{L}(\epsilon_{1})+f_{L}(\epsilon_{2})\right)+\left(f_{R}(\epsilon_{1})\left(1-A\right)^{2}+f_{R}(\epsilon_{2})\left(1+A\right)^{2}\right)\bigg]\\
\alpha_{RF} & = & \frac{\Gamma}{2}\bigg[\left(f_{L}(\epsilon_{1})\left(1-A\right)^{2}+f_{L}(\epsilon_{2})\left(1+A\right)^{2}\right)+B^{2}\left(f_{R}(\epsilon_{1})+f_{R}(\epsilon_{2})\right)\bigg]\label{eq:alpha}\;.
\end{eqnarray}
For simplicity we assume the leads density of states to be homogenous
and identical $n_{L}(\epsilon)=n_{R}(\epsilon)=n$ as for the coupling
strength $\Gamma^{L}=\Gamma^{R}=\Gamma/n\pi$. The coefficients $\beta_{XY}$
are obtained replacing in the definition above $f_{\nu}(\epsilon)\to1-f_{\nu}(\epsilon)$.
The effective decoherence is written as
\begin{equation}
\mathcal{L}_{C}\left[\bullet\right]=\sum_{\begin{array}{c}
{\scriptscriptstyle XY=\{LO,LF,RO,RF\}}\\
{\scriptscriptstyle AB=\{RO,RF,LO,LF\}}
\end{array}}\bigg[\alpha_{XY}^{c}\left(\hat{L}_{XY}\bullet\hat{L}_{AB}^{\dagger}-\frac{1}{2}\left\{ \hat{L}_{AB}^{\dagger}\hat{L}_{XY},\bullet\right\} \right)+\beta_{XY}^{c}\left(\hat{L}_{AB}^{\dagger}\bullet\hat{L}_{XY}-\frac{1}{2}\left\{ \hat{L}_{XY}\hat{L}_{AB}^{\dagger},\bullet\right\} \right)\bigg]\label{eq:lc}\;.
\end{equation}
It is worth to notice that those dissipator acting on the coherence
are not of a dephasing form because they change the energy of the
system. They give rise to or suppress coherences between the 2 intermediate
states through absorption or emission of quanta of energy with the
leads. The coupling strength attached to those channels are such as
\begin{eqnarray}
\alpha_{O}^{c}=\alpha_{LO}^{c}=\alpha_{RO}^{c} & = & \frac{\Gamma}{2}\bigg[\left(1-A\right)\left(f_{L}(\epsilon_{1})B^{2}+f_{R}(\epsilon_{1})\left(1-A\right)^{2}\right)-\left(1+A\right)\left(f_{L}(\epsilon_{2})B^{2}+f_{R}(\epsilon_{2})\left(1+A\right)^{2}\right)\bigg]\\
\alpha_{F}^{c}=\alpha_{LF}^{c}=\alpha_{RF}^{c} & = & \frac{\Gamma}{2}\bigg[\left(1-A\right)\left(f_{L}(\epsilon_{1})\left(1-A\right)^{2}+f_{R}(\epsilon_{1})B^{2}\right)-\left(1+A\right)\left(f_{L}(\epsilon_{2})\left(1+A\right)^{2}+f_{R}(\epsilon_{2})B^{2}\right)\bigg]\;.\label{eq:alphac}
\end{eqnarray}
In order to retrieve the exchange statistics we define as done in the text
(section \ref{sub:Exchange-statistics}) a counting process $K$ which
is related to a biased contribution of the evolution $\mathcal{L}_{s}$
where
\begin{eqnarray}
\mathcal{L}_{s}[\bullet] & = & \sum_{b=\{O,F\}}\sum_{a=\{L,R\}}\Bigg[\left(e^{\left(-1\right)^{b}s}-1\right)\Bigg(\alpha_{b}^{Rc}\vert a\rangle\langle b\vert\bullet\vert b\rangle\langle\bar{a}\vert+\alpha_{ab}^{R}\vert10\rangle\langle b\vert\bullet\vert b\rangle\langle10\vert\Bigg)\nonumber \\
 &  & +\left(e^{\left(-1\right)^{b+1}s}-1\right)\Bigg(\beta_{b}^{Rc}\vert b\rangle\langle a\vert\bullet\vert\bar{a}\rangle\langle b\vert+\beta_{ab}^{R}\vert b\rangle\langle a\vert\bullet\vert a\rangle\langle b\vert\Bigg)\Bigg]\label{eq:ls}
\end{eqnarray}
with $\bar{a}$ is the complementary of $a$ ($R$ for $L$ and reciprocally).

\subsection*{Appendix 2: Superoperator of the effective electronic dynamics \label{sub:superop}}

Here we provide the explicit form of the superoperator responsible for the effective open-system dynamics of the electronic system discussed in Sec.~\ref{sec:Wire-dynamic}. With the definitions of the previous Appendix, we have
\begin{equation}
\boldsymbol{\mathcal{W}}=\begin{pmatrix}-\left(\alpha_{0,0}+\alpha_{0,2}\right) & \beta_{RO} & \beta_{LO} & 0 & \beta_{O}^{c} & \beta_{O}^{c}\\
\alpha_{RO} & -\left(\beta_{RF}+\beta_{RO}\right)-\Gamma_{\text{v}} & \Gamma_{\text{v}} & \alpha_{RF} & -\frac{1}{2}\left(\beta_{O}^{c}+\beta_{F}^{c}\right) & -\frac{1}{2}\left(\beta_{O}^{c}+\beta_{F}^{c}\right)\\
\alpha_{LO} & \Gamma_{\text{v}} & -\left(\beta_{LO}+\beta_{LF}\right)-\Gamma_{\text{v}} & \alpha_{LF} & -\frac{1}{2}\left(\beta_{O}^{c}+\beta_{F}^{c}\right) & -\frac{1}{2}\left(\beta_{O}^{c}+\beta_{F}^{c}\right)\\
0 & \beta_{RF} & \beta_{LF} & -\left(\alpha_{3,0}+\alpha_{3,2}\right) & \beta_{F}^{c} & \beta_{F}^{c}\\
\alpha_{O}^{c} & -\frac{1}{2}\left(\beta_{O}^{c}+\beta_{F}^{c}\right) & -\frac{1}{2}\left(\beta_{O}^{c}+\beta_{F}^{c}\right) & \alpha_{F}^{c} & -i\Delta-\Gamma_{\text{v}}-\frac{1}{2}\beta & \Gamma_{\text{v}}\\
\alpha_{O}^{c} & -\frac{1}{2}\left(\beta_{O}^{c}+\beta_{F}^{c}\right) & -\frac{1}{2}\left(\beta_{O}^{c}+\beta_{F}^{c}\right) & \alpha_{F}^{c} & \Gamma_{\text{v}} & i\Delta-\Gamma_{\text{v}}-\frac{1}{2}\beta
\end{pmatrix}\label{eq:ddd}
\end{equation}
with $\beta=\beta_{LO}+\beta_{LF}+\beta_{RF}+\beta_{RO}$ and $\Delta=\epsilon_{R}-\epsilon_{L}$. The explicit definition of
each coefficient is given in Appendix \ref{sub:Dissipation-of-the}. Notice that each coefficient $\alpha$ and $\beta$ can be easily rewritten in terms of leads contribution such as $\alpha_{XY}=\alpha_{XY}^{R}+\alpha_{XY}^{L}$. 

The effective dissipator of the biased master equation resulting form the inclusion of counting processes has the following matrix representation, instead
\begin{equation}
\boldsymbol{\mathcal{L}}_{s}=\begin{pmatrix}0 & \beta_{RO}^{R}\left(e^{-s}-1\right) & \beta_{LO}^{R}\left(e^{-s}-1\right) & 0 & \beta_{O}^{Rc}\left(e^{-s}-1\right) & \beta_{O}^{Rc}\left(e^{-s}-1\right)\\
\alpha_{RO}^{R}\left(e^{s}-1\right) & 0 & 0 & \alpha_{RF}^{R}\left(e^{-s}-1\right) & 0 & 0\\
\alpha_{LO}^{R}\left(e^{s}-1\right) & 0 & 0 & \alpha_{LF}^{R}\left(e^{-s}-1\right) & 0 & 0\\
0 & \beta_{RF}^{R}\left(e^{s}-1\right) & \beta_{LF}^{R}\left(e^{s}-1\right) & 0 & \beta_{F}^{Rc}\left(e^{s}-1\right) & \beta_{F}^{Rc}\left(e^{s}-1\right)\\
\alpha_{O}^{Rc}\left(e^{s}-1\right) & 0 & 0 & \alpha_{F}^{Rc}\left(e^{-s}-1\right) & 0 & 0\\
\alpha_{O}^{Rc}\left(e^{s}-1\right) & 0 & 0 & \alpha_{F}^{Rc}\left(e^{-s}-1\right) & 0 & 0
\end{pmatrix}\label{eq:ddd-1}.
\end{equation}
\end{widetext}

\end{document}